\documentclass[aps, pra, reprint, noshowpacs, superscriptaddress,nofootinbib,longbibliography,floatfix]{revtex4-2} \newcommand{\abstractbox}[1]{#1}

\usepackage{graphicx}
\usepackage{array}
\usepackage{url}
\usepackage{multirow}
\usepackage{soul}
\usepackage{xcolor}

\usepackage{mathtools,amssymb}
\usepackage{tikz}
\usepackage{tabularx}
\usepackage{enumitem}
\usepackage{censor}
\usepackage{svg}
\usepackage{calc}

\newcolumntype{L}{>{\raggedright\arraybackslash}X}
\newcolumntype{C}{>{\centering\arraybackslash}X}
\newcolumntype{s}{>{\hsize=48mm}L}

\newcommand{\note}[1]{}
\definecolor{myred}{rgb}{.85,0,0}

\begin{document}
\title{Overview of couplet scoring in content-focused physics assessments}

\author{Michael Vignal}
\affiliation{JILA, National Institute of Standards and Technology and the University of Colorado, Boulder, CO 80309, USA}
\affiliation{Department of Physics, University of Colorado, 390 UCB, Boulder, CO 80309, USA}
\affiliation{Department of Physics, Willamette University, 900 State Street Salem, OR 97301, USA}

\author{Gayle Geschwind}
\affiliation{JILA, National Institute of Standards and Technology and the University of Colorado, Boulder, CO 80309, USA}
\affiliation{Department of Physics, University of Colorado, 390 UCB, Boulder, CO 80309, USA}

\author{Marcos D. Caballero}
\affiliation{Department of Physics \& Astronomy and CREATE for STEM Institute, Michigan State University, East Lansing, Michigan 48824, USA}
\affiliation{Department of Computational Mathematics, Science, \& Engineering, Michigan State University, East Lansing, Michigan 48824, USA}
\affiliation{Department of Physics and Center for Computing in Science Education, University of Oslo, 0315 Oslo, Norway}

\author{H. J. Lewandowski}
\affiliation{JILA, National Institute of Standards and Technology and the University of Colorado, Boulder, CO 80309, USA}
\affiliation{Department of Physics, University of Colorado, 390 UCB, Boulder, CO 80309, USA}

\begin{abstract}
\abstractbox{Content-focused research-based assessment instruments typically use items (i.e., questions) as the unit of assessment for scoring, reporting, and validation.  Couplet scoring employs an alternative unit of assessment called a couplet, which is essentially an item viewed and scored through the lens of a specific assessment objective. With couplet scoring, a single item may have more than one assessment objective and therefore more than one couplet and thus more than one score. We outline the components of traditional item scoring, discuss couplet scoring and its benefits, and use both a recently developed content research-based assessment instrument and an existing one to ground our discussion.}
\end{abstract}

\maketitle

\section{Introduction}

Research-based assessment instruments (RBAIs) help educators (including instructors and education researchers) make informed pedagogical and curricular decisions~\citep{aera_standards_nodate,madsen_best_2017,madsen_resource_2017,maric_measurement_2023}. These instruments can be surveys, questionnaires, and other tools used to gather information on student beliefs, experiences, proficiencies, and other aspects of education that are of interest to educators. Unlike summative assessments that typically evaluate individual students, RBAIs are intended to identify trends in populations of students~\citep{engelhardt_introduction_2009,madsen_best_2017}. Here, we focus on primarily RBAIs that measure student proficiency (skills and knowledge) in specific content areas, which we refer to as \textit{content RBAIs}. 

Recently, we created a physics content RBAI for measurement uncertainty~\citep{vignal_survey_2023}, employing assessment objectives (AOs)~\citep{vignal_affordances_2022} throughout the development process. AOs are statements (similar in structure to learning objectives~\citep{anderson_taxonomy_2001}) about the content the instrument aims to assess. For our RBAI, these AOs are integral to the interpretation, scoring, and reporting of student responses, as each item is designed to align with one or more AO.

Our use of AOs supported developing an instrument that aligned with our assessment priorities. Indeed, the usefulness of RBAIs depends, in large part, on the degree to which an instrument measures what it purports to measure and how meaningfully these measures are reported to implementers~\citep{hemphill_measurement_1950,rovinelli_use_1977,nunnally_psychometric_1994,lindell_establishing_2013,crocker_introduction_2008,miller_measurement_2009,engelhardt_introduction_2009,maric_measurement_2023}. Typically, these measures are item scores, and they are often reported individually or as an overall assessment score~\citep{miller_measurement_2009,engelhardt_introduction_2009}.

Here, we formalize our AO-aligned scoring paradigm for content RBAIs called \textit{couplet scoring}, where a couplet is a scorable item-AO pair. In this paradigm, it is couplet scores, rather than item scores, that serve as the unit of assessment for reporting student proficiencies and validating the instrument. We posit that couplet scoring offers a number of benefits as compared to dichotomous item scoring, in which each item has only one correct answer and the assessment score is generally the sum of the item scores. 

The instrument for which we developed couplet scoring is the Survey of Physics Reasoning on Uncertainty Concepts in Experiments (SPRUCE)~\citep{vignal_survey_2023}. SPRUCE was developed in parallel with couplet scoring and is used throughout this paper to demonstrate how this scoring paradigm might be applied to content RBAIs. However, the focus of this paper is couplet scoring and not SPRUCE itself, and so we also incorporate examples from ongoing work to generate a \textit{post hoc} couplet scoring scheme for the Force Concept Inventory (FCI).

The work presented here has the following research goals:

\begin{enumerate}
    \item[RG1] Describe couplet scoring and explore its benefits and limitations; and
    \item[RG2] Demonstrate how couplet scoring can be employed in content RBAI development.
\end{enumerate}

In Sec.~\ref{sec traditional}, we provide background on assessments and scoring schemes. Sec.~\ref{sec scoring by couplet} then discusses our new scoring paradigm, and we outline some of its benefits in Sec.~\ref{sec affordances}. Details of the implementation of couplet scoring are shared in Sec.~\ref{sec implementation}. A summary, and discussion of future work, and discussion of possible implications for other types of evaluation are presented in Sec.~\ref{summary}.

\section{Background}
\label{sec traditional}

Research-based assessment instruments (RBAIs) are tools used by educators and researchers to gather information from students about teaching, learning, student experiences, and other aspects of education to inform curricular and pedagogical decisions. These instruments are ``developed based on research into student thinking...[to] provide a standardized assessment of teaching and learning'' when administered to students~\citep{madsen_resource_2017}. Importantly, these instruments are not intended to help educators evaluate individual students or assign grades~\citep{engelhardt_introduction_2009,madsen_best_2017}.

In physics, most RBAIs are designed to measure student proficiencies in specific content areas, such as in mechanics~\citep{hestenes_force_1992,thornton_assessing_1998}, 
electricity and magnetism~\citep{maloney_surveying_2001,chabay_qualitative_1997,wilcox_coupled_2014}, 
quantum mechanics~\citep{sadaghiani_quantum_2015}, 
thermodynamics~\citep{rainey_designing_2020}, 
or laboratory settings~\citep{campbell_teaching_2005,walsh_quantifying_2019,day_development_2011,dounas-frazer_characterizing_2018,vignal_survey_2023}.
These \textit{content RBAIs} (which include concept inventories~\citep{madsen_best_2017} and conceptual assessment instruments~\citep{lindell_establishing_2013}) have proven valuable in identifying instructional weaknesses~\citep{madsen_best_2017,
mazur_farewell_2009
} and evaluating the effectiveness of instructional changes~\citep{kohlmyer_tale_2009,goldhaber_transforming_2009,caballero_comparing_2012,madsen_best_2017,pollard_impact_2020} in physics education. These and other instruments can be found on the PhysPort website~\citep{mckagan_physport_nodate}.


RBAIs can employ a wide variety of item types, including both selected-response formats (such as multiple choice) and constructed-response formats (such as written explanations or providing a number). Selected-response format items, especially multiple-choice items, are generally developed such that one response is considered the correct choice (typically determined by evaluating alignment with expert responses~\citep{engelhardt_introduction_2009}) and other responses are considered incorrect. This convention supports the inclusion of multiple-choice items as opposed to multiple-response (sometimes called multiple-select or multiple-choice-multiple-response~\citep{smith_analyzing_2022,white_brahmia_physics_2021}) and coupled multiple response items~\citep{wilcox_coupled_2014}, as single-correct-answer items require minimally complex scoring mechanisms. This scoring approach is often described as being the most objective~\citep{nunnally_psychometric_1994,engelhardt_introduction_2009} and it generates scores that work well with validation algorithms~\citep{nunnally_psychometric_1994}. Even when using item formats other than multiple-choice, having a single correct answer is common and aligns with instructor and student expectations around assessment.

Many instruments, however, go beyond the strict ``one correct answer per item'' scoring model, either by design or as the result of post-development modifications. Many instruments employ various two-tier questions that reward consistency even if the first "tier" of the item is answered incorrectly, with perhaps coupled multiple-response items (as discussed in Refs.~\citep{wilcox_coupled_2014},~\citep{rainey_designing_2020} and~\citep{rainey_validation_2022}) representing the greatest departure from conventional assessment items. The Brief Energy and Magnetism Survey scoring scheme was modified so that several items have scores that depend on the response to that item, as well as to previous, related items~\citep{chabay_qualitative_1997}. The developers of the Force and Motion Conceptual Evaluation~\citep{thornton_assessing_1998} advocate for an alternative scoring scheme that includes consolidating 3 different groups of 3 questions and awarding 2 points if all the items in a group are answered correctly~\citep{thornton_comparing_2009}. 

To capture and report more information from existing RBAIs, researchers have worked to identify sub-scales related to different concepts covered on the RBAI~\citep{miller_measurement_2009,stewart_multidimensional_2018,hansen_multidimensional_2021}, with sub-scale mean scores being reported in addition to an overall mean score. However, such sub-scale analyses are typically time and labor-intensive to develop, external to the  report provided to instructors, and not considered in the design and validation of the instrument~\citep{miller_measurement_2009}. For example, Stewart et al.~\citep{stewart_multidimensional_2018} and Hansen and Stewart~\citep{hansen_multidimensional_2021} found the FCI and another popular physics content RBAI had a ``lack of coherent sub-scales'', limiting the usefulness of such post hoc sub-scale approaches.

Additionally, work has been done to learn about student reasoning and proficiencies from their incorrect answers via weighted scoring~\citep{smith_motivations_2022,brewe_using_2016}\note{IRT nominal response model, module analysis}, with the idea being that not all distractors represent the same lack of knowledge or skills. While much of this work is done post development, some developers have worked to encode some information about the quality of incorrect answers through partial credit scoring schemes, where some distractors, rather than being worth no points, are worth a fraction of the points that are given for the correct answer.  While this scoring model can be more sensitive in measuring student proficiencies, the exact fraction of a point earned for these answers is subjective and can restrict which validation algorithms one can use, removing one of the advantages gained by having single-correct items. Additionally, the nuance of exactly what part of a response earned a student credit is difficult to capture and convey in an overall assessment scores or even in the item score.

Finally, a related method, called choice modeling or choice analysis, has been used by researchers in economics, marketing, and transportation to explore choices made by survey respondents based on both selected and not selected options.  Choice-level analyses can extract information beyond the final result, and are used to understand the opinions and rationale behind peoples' decisions~\citep{mcfadden_conditional_1972, louviere_stated_2000, ben-akiva_discrete_1985}. However, this context is well outside of the sphere of RBAIs and physics education, being used to understand priorities and preferences rather than skills and proficiencies.

Here, we present another model of scoring an assessment, called couplet-scoring, that goes beyond scoring an item as either correct  or incorrect and allows for additional information to be reported about student reasoning along several constructs. We will discuss how consideration of each of the above ideas contributes to the development of this scoring scheme.

\section{Couplet Scoring for Research-Based Assessment Instruments}
\label{sec scoring by couplet}

In this section, we present our new assessment scoring paradigm, couplet scoring. Central to couplet scoring is the use of assessment objectives (AOs), which we introduced previously~\citep{vignal_affordances_2022,vignal_survey_2023} and summarize below. 


\subsection{Assessment Objectives}

AOs are ``concise, specific articulations of measurable desired student performances regarding concepts and/or practices targeted by the assessment''~\citep{vignal_affordances_2022}. In the language of assessment development, AOs are the constructs the assessment aims to measure, and they are articulated explicitly as objectives. Table~\ref{tab example AOs} includes some example AOs from SPRUCE and several theoretical principles developed by Stewart et al. for the FCI~\citep{stewart_multidimensional_2018}, which we employ as preliminary AOs in this paper\footnote{These principles were developed as part of a multidimensional IRT analysis of the FCI and not as a set of AOs. However, they span a similar set of ideas as would AOs, and thus are adequate stand-ins for AOs for the example presented in this paper.} to illustrate what couplet scoring might look like with the FCI.

\begin{table}[t!]
    \caption{Examples of AOs from SPRUCE and theoretical principles from the FCI that serve as the item constructs in the examples below.}
    \label{tab example AOs}
    \centering
    \begin{tabularx}{\linewidth}{X}\hline\hline
        \multicolumn{1}{c}{SPRUCE AO examples} 
        \\\hline
        \begin{itemize}
            \item [S2] Identify actions that might improve precision
            \item [S3] Identify actions that might improve accuracy
            \item [H1] Propagate uncertainties using formulas
            \item [H2] Report results with uncertainties with correct significant digits
            \item [D1] Articulate why it is important to take several measurements during experimentation
            \item [D5] Determine if two measurements (with uncertainty) agree with each other
        \end{itemize}
        \\\hline
        \multicolumn{1}{c}{FCI theoretical principle examples~\citep{stewart_multidimensional_2018}} 
        \\\hline
        \begin{itemize}
            \item[C2] Objects moving in a curved trajectory will experience centripetal acceleration
            \item[L2] Newton's 2nd law
            \item[L4] Objects near the earth's surface experience a constant downward force/acceleration of gravity
            \item[F2] An object does not necessarily experience a force in the direction of motion
            \vspace{-\baselineskip}
        \end{itemize}
        \\\hline\hline
    \end{tabularx}
\end{table}

In a previous paper~\citep{vignal_affordances_2022}, we outlined four broad affordances (or benefits) of AOs for instrument development: facilitating incorporating instructor priorities into the instrument, providing a means for evaluating and scoring authentic items, providing a structure for feedback to implementers, and serving as a means for communicating the content of the instrument to potential implementers~\citep{vignal_affordances_2022}. Many of these benefits complement those of couplet scoring, discussed in Sec.~\ref{sec affordances}.

\subsection{Couplet Scoring}
\label{couplet scoring}

Couplet scoring, as the name suggests, is a scoring paradigm in which item-AO couplets (or simply ``couplets'') are scored. This is in contrast to many traditional scoring paradigms, in which each item is scored once.

\begin{figure}[ht]
    \centering
    \includegraphics[width=0.8\linewidth]{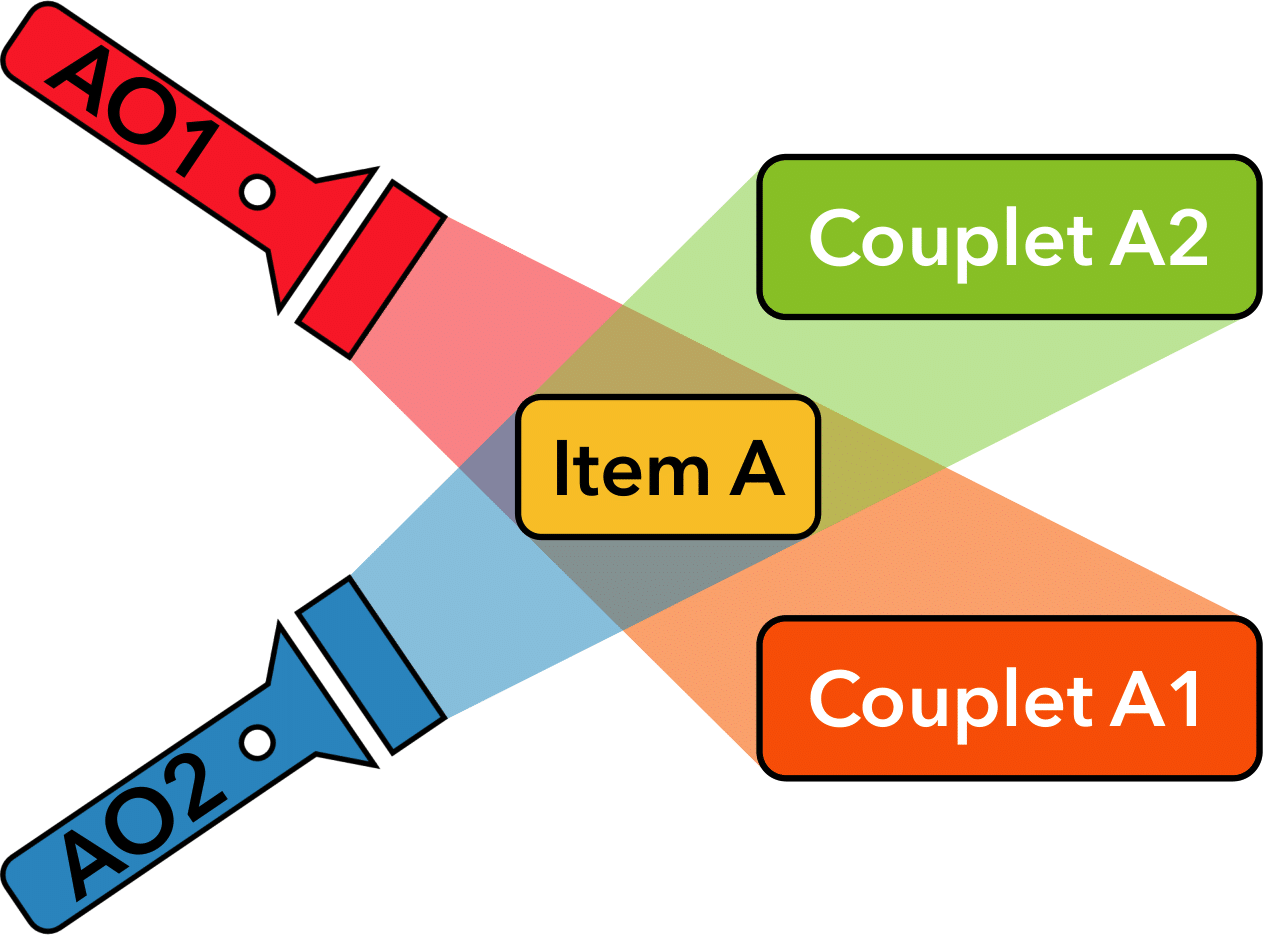}
    \caption{A simple graphic depicting how two AOs, represented as flashlights, can ``illuminate'' the same item and yet produce independent couplets.}
    \label{coupletcartoon}
\end{figure}

Conceptually, a couplet is an assessment item viewed and scored in light of a particular AO. Multi-AO items have a couplet for each AO, as depicted in Fig.~\ref{coupletcartoon} for an item that has two AOs and therefore two couplets. Each of these couplets is scored by considering only that couplet's AO. 

\subsubsection{Couplet scoring example - a simple model}

\begin{figure*}[ht]
    \centering
    \includegraphics[width=0.8\linewidth]{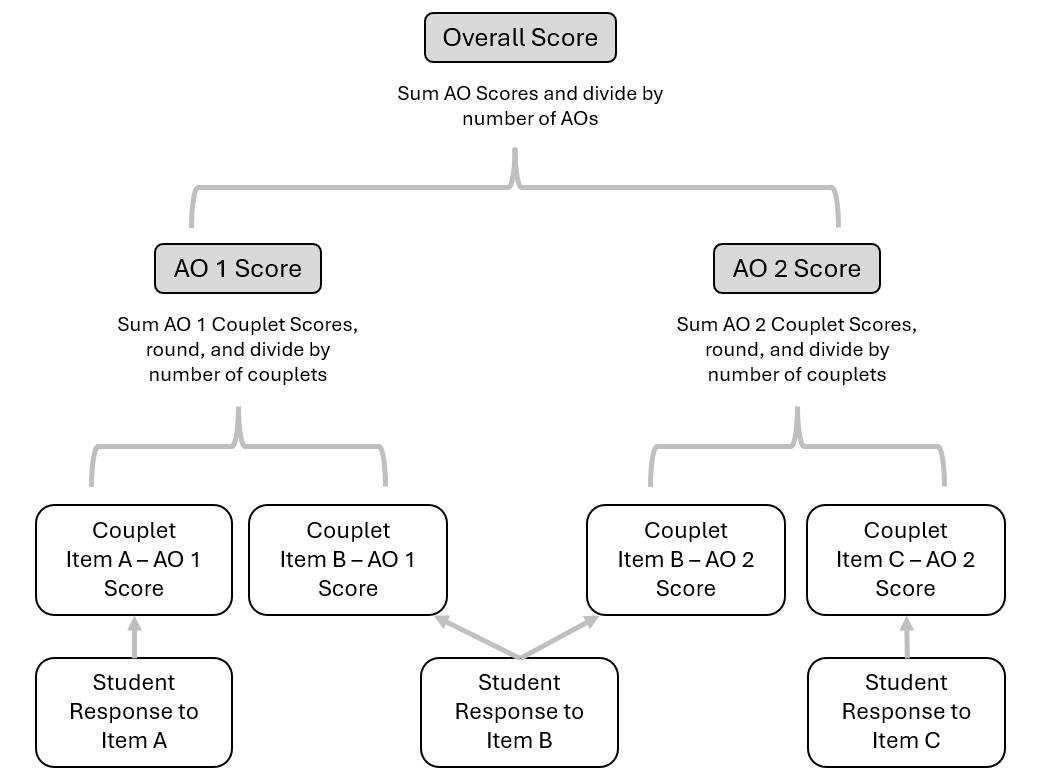}
    \caption{Flowchart showing how a student's responses to the model instrument items become couplet, AO, and finally overall scores. Generally, student response are scored along relevant AOs. These couplet scores are summed, rounded to the nearest integer, and normalized to 1 for each AO. Finally, these AO scores are summed and normalized to 1 in order to create an overall score that equally weights each AO. Gray shading indicates numbers reported to instructors and used in research analysis.}
    \label{flowchart}
\end{figure*}

Similar to how the image in Fig.~\ref{coupletcartoon} uses a simple graphic to illustrate how AOs and items combine to produce couplets, we now use a simple model of an instrument (comprised of only three multiple choice items and two AOs) to illustrate how couplet scoring is implemented and how the AO scores and the overall score (for one student) can be calculated. Table~\ref{table toyscoring} shows the AO scores associated with each possible answer on this instrument. Figure~\ref{flowchart} shows how to use student responses to the items in the model to compute couplet scores, AO scores, and, finally, an overall score.

\begin{table}[ht]
    \centering
    \caption{Couplet scoring scheme for a sample instrument comprised of three items. Note that Item A does not probe AO2 and Item C does not probe AO1 in this example, similar to a real-world application where not every item will probe every AO.}
    \label{table toyscoring}
    \begin{tabular}{ccc}\hline\hline
        Answer Option & \multicolumn{2}{c}{Score}\\\hline
        \multicolumn{3}{c}{Item A}\\
        & A01 & A02 \\\hline
        a & 1 & -  \\
        b & 1 & -  \\
        c & 0 & -  \\
        d & 0 & -  \\
        e & 0 & -  \\
        f & 0 & - \\\hline
        \multicolumn{3}{c}{Item B}\\
        & A01 & A02 \\\hline
        a & 0 & 0  \\
        b & 1 & 1  \\
        c & 0 & 1  \\
        d & 1 & 0  \\
        e & 1 & 1 \\ \hline

        \multicolumn{3}{c}{Item C}\\
        & A01 & A02 \\\hline
        a & - & 1  \\
        b & - & 0  \\
        c & - & 1  \\
        d & - & 0  \\\hline
        
        \hline\hline
    \end{tabular}
\end{table}

Since many answer options for items on RBAIs are developed to be tempting distractors (i.e., results that may be correct in some ways but incorrect in others), it is possible (and, for SPRUCE, fairly common) for a couplet to award the maximum number of points to several different possible responses, even for selected-response items such as multiple-choice items. 

To calculate an AO score for a student, each item that addresses that AO is scored to form a couplet score (See Fig.~\ref{flowchart}). The couplet scores for that AO are summed together, rounded to the nearest integer, and divided by the number of couplets for that AO. This produces an AO score with a minimum of 0 and a maximum of 1. 

While the specifics of calculating scores should reflect the priorities of the developers and instructors, here we discuss an example based on the method used to calculate scores for SPRUCE. In Tab.~\ref{table toyscoring}, we can see a student who selected option b on all three of items A, B, and C would have AO scores of 1 on AO 1 (one point for couplet AO 1 - Item A, one point on couplet AO 1 - Item B, and item C is not scored in this couplet) and 0.5 for AO2 (item A is not scored for this couplet, one point on couplet AO 2 - Item B, and no points on couplet AO2 - Item C). Similarly, a second student who selected option c on item A, option b on item B and option d on item C would have AO scores of 0.5 on AOs 1 and 2.

Once all the AO scores are calculated for a student, they can then be summed together and divided by the number of AOs to create an overall score, which has a minimum of 0 and a maximum of 1, giving  an average of the AO scores (as opposed to an average of couplet scores), refleting a desire to equally weigh AOs.  The average score for all students for each AO, as well as the average overall score, are all reported to instructors and can be used for answering research questions. While the overall score is reported to instructors to allow them to get a general sense of student performance on the topic covered by the RBAI (such as Newtonian mechanics or measurement uncertainty), the finer-grained AO scores can help instructors focus their efforts to improve the course.

The remainder of this section explores examples of couplet scoring from two RBAIs, SPRUCE and the FCI. Despite differences in these instruments, these examples show how specific items can be scored using a couplet scoring scheme to produce multiple couplet scores that contribute to various instrument AO scores.

\subsubsection{Couplet scoring example - the Survey of Physics Reasoning on Uncertainty Concepts in Experiments} 

\begin{figure}[h]
    \centering
    \includegraphics[width=.9\linewidth]{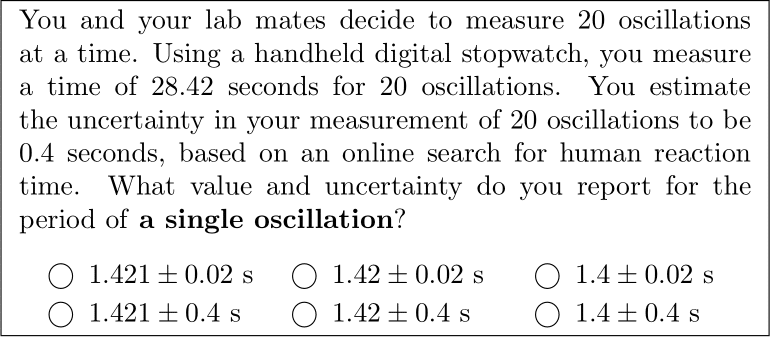}
    \caption{SPRUCE item 3.3, in which students are attempting to determine the period of oscillation for a mass hanging vertically from a spring. This item has two AOs, H1 and H2. We have modified the numbers in this example to protect test security.}
    \label{fig 33}
\end{figure}

An example of a multiple-choice item scored using couplet scoring, item 3.3 from SPRUCE (shown in Fig.~\ref{fig 33}) tasks students with determining the period of oscillation for a mass hanging vertically from a spring. This item has two AOs, which we refer to as ``3.3~H1'' and ``3.3~H2''.

\begin{itemize}
    \item [H1] Propagate uncertainties using formulas
    \item [H2] Report results with uncertainties with correct significant digits
\end{itemize}

\noindent The proficiencies represented by AOs H1 and H2 both impact the response a student would provide on item~3.3, but they are conceptually independent of one another: thus it is straightforward to create an item and scoring scheme that independently determines if students correctly propagate uncertainty (H1) and if they report their value and uncertainty with correct significant digits (H2).

For couplet 3.3 H1, applying the appropriate uncertainty propagation formula simplifies to dividing the uncertainty in the time for 20 oscillations by 20, yielding a value of $\pm 0.02$~s as the uncertainty in the period of a single oscillation. This value appears in three of the answer options, and so the selection of any of these three responses awards a full point for couplet 3.3~H1. Similarly, two of the six answer options are presented with appropriate numbers of significant digits for the value of the period (based on the value of the uncertainty) and thus receive full credit for couplet 3.3 H2. 

\begin{table}[!t]
    \centering
    \caption{Example scoring for couplets of item~3.3}
    \label{table 33scoring}
    \begin{tabular}{llcc}\hline\hline
        \multicolumn{2}{l}{\multirow{2}{*}{Answer Option}} & \multicolumn{2}{c}{Score}\\
        & & H1 & H2                  \\\hline
        A & $1.421 \pm 0.02$ s  & 1 & 0 \\
        B & $1.421 \pm 0.4$ s   & 0 & 0 \\
        C & $1.42 \pm 0.02$ s   & 1 & 1 \\
        D & $1.42 \pm 0.4$ s    & 0 & 0 \\
        E & $1.4 \pm 0.02$ s    & 1 & 0 \\
        F & $1.4 \pm 0.4$ s     & 0 & 1 \\\hline\hline
    \end{tabular}
\end{table}

SPRUCE item 3.3's couplets may seem sufficiently independent that it would be straightforward to split the item into two independent items, one for each AO. However, such a separation will not meaningfully measure student proficiencies with H2 unless the response of the item targeting H2 is coordinated with the response of couplet 3.3~H1, as proper reporting requires coordinating between the significant digits of the result and the uncertainty. Additionally, splitting the item such that one item targets only H2 would likely involve prompting students to use proper significant digits (which SPRUCE does do with other couplets), rather than what the current item does, which is measure if students use proper significant digits without prompting. In these ways, couplet scoring allows these two AOs to be targeted in ways that two separate, traditionally scored items could not.

\subsubsection{Couplet scoring example - the Force Concept Inventory} We now consider item 18 from the FCI~\citep{hestenes_force_1992} and present one potential couplet scoring scheme for this item. The purpose of this example is to demonstrate how couplet scoring may (i) yield more actionable feedback than dichotomous item scoring; (ii) identify and inform potential modifications to items; and (iii) support or improve upon various instrument analyses. We are not arguing that a couplet scoring scheme should be developed for the FCI---though ongoing work is investigating such a scoring scheme---and we will discuss the limitations of applying a couplet scoring scheme \textit{post hoc} to an instrument designed with traditional scoring in mind.

The FCI continues to be extensively deployed and studied (e.g.,~\citep{brewe_using_2016,stewart_multidimensional_2018,eaton_generating_2019}). To imagine a couplet scoring scheme for item 18 of the FCI, we repurpose Stewart et al.'s \citep{stewart_multidimensional_2018} principles---developed based on both theory and refined using multidimensional item response theory (MIRT)---as our AOs. Principles are ``fundamental reasoning steps in the solution of the item''~\cite{hansen_multidimensional_2021}, rather than statements about what the instrument can measure, but, as we will see, the principles can serve as a reasonable starting place for developing AOs.

\begin{figure}[t]
    \centering
    \includegraphics[width=\linewidth]{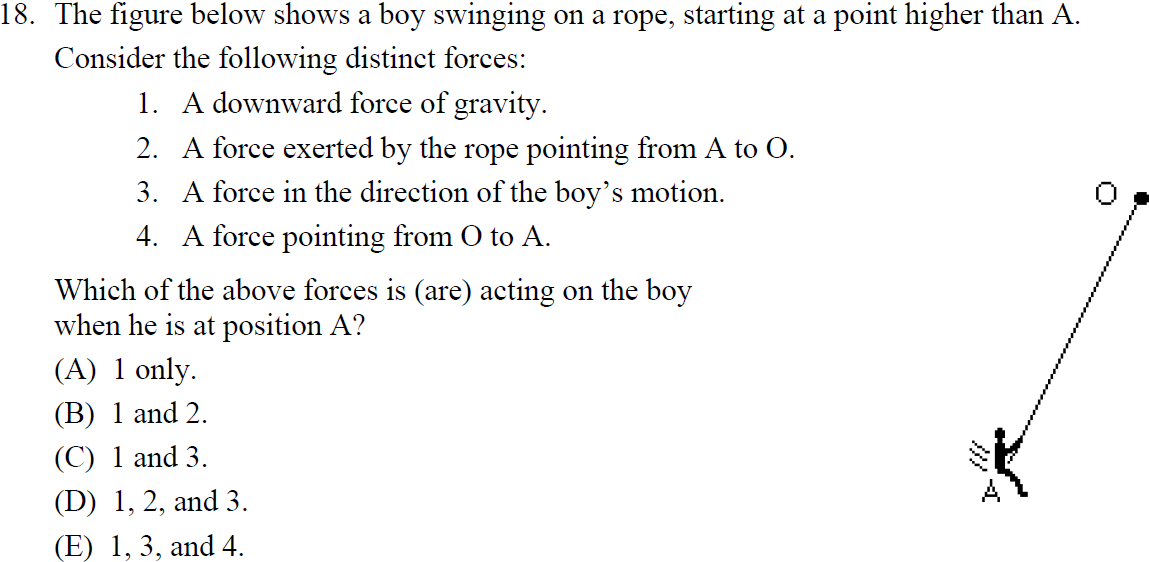}
    \caption{Item 18 from the Force Concept Inventory.}
    \label{fciimage}
\end{figure}

Item 18 of the FCI, shown in Fig.~\ref{fciimage}, asks students to consider which of four possible forces are acting on a boy swinging on a rope: gravity, a tension force towards the boy, a force in the direction of motion, and/or a tension force away from the boy. With four possible forces, there are theoretically $2^4=16$ possible responses, of which students are asked to choose one amongst five.

Stewart et al. identified four theoretical principles for this item, all of which were retained in their optimal employed MIRT model. These principles are:
\begin{enumerate}
    \item[C2] Objects moving in a curved trajectory will experience centripetal acceleration
    \item[L2] Newton's 2nd law
    \item[L4] Objects near the earth's surface experience a constant downward force/acceleration of gravity
    \item[F2] An object does not necessarily experience a force in the direction of motion
\end{enumerate}

With these principles serving as our AOs, we consider each of the four forces described in the item independently in terms of each AO and determine how the selection of each force indicates proficiency with each AO. Incorrect, neutral, partially correct, and correct applications of the AOs are respectively represented by numbers that are negative, equal to zero, between zero and one, and equal to one, as shown in Table~\ref{fcischemeparts}.

\begin{table}[!ht]
\caption{Independent scoring, by AO, of each force that is represented in the answer options of item 18 on the Force Concept Inventory. These scorings will contribute to the couplet scores for each answer option via a simple algorithm that sums the scores and then applies a floor of 0 and a ceiling of 1.}
    \centering
    \begin{tabular}{lcccc}\hline\hline
        & \multicolumn{4}{c}{Principle/AO}\\
        Force & C2 & L2 & L4 & F2  \\\hline
         1 (gravity) & 0 & 0.5 & 1 & 1 \\
         2 (radial in) &1 & 0.5 & 0 & 1 \\
         3 (along motion) & 0 & 0 & 0 & -10 \\
         4 (radial out) & -1 & 0 & 0 & 1\\\hline\hline
    \end{tabular}
    
    \label{fcischemeparts}
\end{table}
\noindent As can be seen in Table~\ref{fcischemeparts}, for C2, zero points are awarded for forces~1 and 3 (because they do not relate to centripetal acceleration), one point is awarded for force~2 (because the force is radial), and one point is subtracted for force~4 (because it is radial but not a force acting on the boy). For AO L2, half a point is awarded for each force~1 and 2 because, together, they comprise the net force acting on the boy. For AO L4, a point is awarded for force~1 (and no other forces) because force~1 is the force of gravity. Finally, for F2, a point is awarded for any force that does not point in the direction of motion, and ten points are subtracted for the selection of force~3 to negate any points that could have been earned for this AO by selecting other forces.

These values can be summed for each selected response for the item, with a floor of zero and a ceiling of one being applied to produce a couplet score. Such a scoring approach can be applied to all 16 possible combinations or forces, including the five available on the FCI (shown in in Table~\ref{fcischeme}).

\begin{table}[tb]
 \caption{Preliminary couplet scores for item 18 on the Force Concept Inventory.}
    \centering
    \begin{tabular}{lcccc}\hline\hline
        & \multicolumn{4}{c}{Principle/AO}\\
        Selection & C2 & L2 & L4 & F2  \\\hline
         (A) 1 only. & 0 & 0.5 & 1 & 1 \\
         (B) 1 and 2. & 1 & 1 & 1 & 1 \\
         (C) 1 and 3. & 0 & 0.5 & 1 & 0 \\
         (D) 1, 2, and 3. & 1 & 1 & 1 & 0\\
         (E) 1, 3, and 4. &  0 & 0.5 & 1 & 0 \\\hline\hline
    \end{tabular}
   
    \label{fcischeme}
\end{table}

This scoring scheme produces five different results for the five different answer options, while dichotomous scoring would yield the same score for all distractors, which we believe represents a loss of potentially actionable information. While this scoring scheme has not been piloted in interviews to develop robust evidentiary arguments, we can, as an exercise, collapse the four couplet scores into a single normalized score for each answer option to approximate a partial-credit scoring scheme. Doing so creates a hierarchy of responses $B>D>A>E=C$ which is quite similar to Eaton et al.'s partial credit hierarchy of $B>D=A>E>C$ that was obtained from a two-parameter logistic nominal response model using polytomous item response theory~\citep{eaton_generating_2019}. While collapsing couplet scoring to a partial-credit model in this way is not aligned with the theoretical foundations of couplet scoring and would no longer be able to produce AO scores for the instrument as a whole, it does provide a convenient reasonableness check for the scoring schemes of individual items.

It is notable that all the available answer options earn a point for L4, as students are not allowed to select an option that does not include a downward gravitational force. Thus, the item (as written) cannot serve as a measurement of student proficiency regarding L4. We would thus remove L4 as an AO for this item, and this process suggests that Stewart et al.'s MIRT analysis of the FCI might reasonably be modified to omit Item 18 from consideration for the factor containing L4. This process also highlights a limitation in applying couplet scoring post development: not being designed to measure multiple AOs in parallel, item 18's answer options were likely selected to remove consideration of gravity as a factor, rather than attempt to measure it.

\section{Couplet Scoring benefits}
\label{sec affordances}

We now expand on the benefits provided by AOs (discussed previously~\citep{vignal_affordances_2022}) to describe some benefits of couplet scoring that are grounded in theory, our own experiences, feedback from users of SPRUCE, and published research. These benefits, summarized in Tab.~\ref{tab affordances}, are organized by whom they primarily benefit, the developer or the user of the instrument.

\begin{table*}[t!]
    \caption{Benefits of couplet scoring and related assessment design features, organized by whom they benefit, the developer or user of the instrument.}
    \label{tab affordances}
    \centering
    \begin{tabularx}{\textwidth}{{>{\hsize=.8\hsize}L>{\hsize=1.3\hsize}L}}\hline\hline
    \multicolumn{1}{c}{Developer Benefit} 
    &\multicolumn{1}{c}{Couplet-Scoring Design Feature}
    \\
    
    Alignment with assessment priorities
    & Alignment is embedded into item development and scoring, as items are developed to be scored by AO, and verification of this alignment is supported by having concise and explicit AOs\\
    
    Construct fidelity
    & Scoring by AO allows for complex, authentic selected-response items with nuanced scoring\\
    
    Scaffolded scoring
    & Developers creating a scoring scheme need only consider one AO at a time\\
    
    Partial credit
    & Reduces, but does not eliminate, need for partial credit scoring by resolving what is “partially” correct and incorrect into different couplets/AO scores\\
    
    Validation
    & Can use many traditional approaches with couplet scores as the unit of assessment\\\hline
    
    \multicolumn{1}{c}{User Benefit} 
    & \multicolumn{1}{c}{Couplet-Scoring Design Feature}
    \\

    Alignment with instructional priorities
    & Instrument AOs (that inform and contextualize scoring) can be directly compared with course learning objectives\\
    
    Data yield
    & Items with multiple AOs yield more data than items with just one AO\\
    
    User experience
    & Indistinguishable from traditional instruments\\
    
    Reporting scores by AO
    & Scores are reported by AO in a manner that is clear and actionable\\\hline\hline    
    \end{tabularx}
\end{table*}

\subsection{Benefits to Developers}

The following benefits are primarily helpful during the design and development of a content RBAI. 

\subsubsection{Alignment with Assessment Priorities}
During instrument development, once items have been created, they undergo an iterative process of piloting and refining, which may change elements of the item such as the item context, the amount of information provided in the prompt, and the available answer options. Each of these changes has the potential to shift the item away from its intended objective. With couplet scoring, the scoring process itself requires that developers revisit the intended objective(s), as proficiency with these objectives, rather than answer-option ``correctness'', is the criteria by which the couplet is scored. Thus, the chance that the final item shifts away from its intended objective(s) is minimized, and if the item does shift, then re-scoring it will require reconsideration of the item's AOs. This built-in reliance of scores on objectives supports alignment at all stages of assessment development, including even before piloting or other data have been collected.

In traditional item scoring, where the scoring schemes do not existentially depend on the item's objective(s), developers must be careful and take additional steps to ensure that the final products align with the intended assessment goals, or else the item may not provide a measure of the targeted and articulated proficiency.

Additionally, having the instrument constructs clearly articulated as AOs facilitates quick and effective verification of alignment between items and constructs by independent expert consultants~\citep{hemphill_measurement_1950,rovinelli_use_1977,engelhardt_introduction_2009}, as was done with SPRUCE~\citep{vignal_survey_2023}. 

\subsubsection{Construct Fidelity}

Though not unique to physics, the synthesis of multiple ideas is often valued in physics instruction and evaluation. This means that items that more authentically depict interesting and relevant physics scenarios often incorporate multiple concepts as a reflection of the interconnected nature of physics~\citep{kimberlin_validity_2008} and thus must often include multiple constructs.

While traditional assessment approaches, by design, typically have assessment items related to only a single construct, couplet scoring allows for more interesting and authentic assessment items that can be scored and interpreted to produce meaningful feedback to users. In this way, these instruments can better serve as appropriate performance assessments~\cite{aera_standards_nodate}. 

As an example, in SPRUCE, the items are all contextualized within four experiments that serve as "testlets", a design feature that allows for non-trivial contexts without overwhelming the student with a new context for each item~\cite{michael_c_rodriguez_selected-responce_2015}. Couplet scoring then facilitates taking advantage of this design to ask rich, authentic questions without the limitation that the items target a single construct.


\subsubsection{Scaffolded Scoring} 

With couplet scoring, the development of a scoring scheme is scaffolded by considering which item responses indicate proficiency in a particular AO. This feature is especially important for scoring schemes with more complex item types, such as multiple-response items and coupled multiple response items.

Anecdotally, when developing couplet scoring schemes for SPRUCE, several members of the research team expressed that this scaffolding made developing a scoring scheme for SPRUCE's coupled multiple-response items feel faster, easier, and less subjective than for coupled multiple response items developed during previous projects.


\subsubsection{Partial Credit}

For selected-response items, students select a response from a list of options that are generally made up of a correct answer and several tempting distractors. These distractors are most effective when they represent an answer that one would arrive at by employing mostly correct reasoning, but also a common misunderstanding or an easy mistake~\citep{engelhardt_introduction_2009}. However, educators often wish to distinguish between different incorrect responses, such as between a response resulting from a simple mistake and one resulting from a fundamental misunderstanding or misapplication of a core concept. As a result, researchers will sometimes employ partial credit scoring schemes (e.g., Ref~\citep{eaton_generating_2019}).

In couplet scoring, each of the lines of reasoning one must employ to obtain the correct answer can often be represented by an AO, and so various distractors may be completely correct in terms of one AO while being incorrect in terms of others. As the item is scored by AO, it is possible for multiple responses to receive full credit for one AO, while not receiving credit for another. This can substantially reduce the need for partial credit, which requires arbitrary weights for partially correct responses. It also better captures and reports the elements of desired reasoning that students employ, since two mostly correct responses will result in meaningfully different couplet scores that do not get obscured by representing the measures of student reasoning with a single number.

For SPRUCE, couplet scoring eliminated the need for partial credit on all items except coupled multiple-response items. In instances where, if using item scoring, we might have considered awarding partial credit for a particular response, we instead were able to award full credit for the one AO and zero credit for another. For example, this can be seen by the couplet scores in table~\ref{table 33scoring} all being zero or one.

\subsubsection{Validation}

The effectiveness of an RBAI is largely contingent on how well it ``measures what it says it measures,'' a property known as validity~\citep{engelhardt_introduction_2009}. The investigation of various types of validity and their metrics is a major topic of scholarship~\citep{hemphill_measurement_1950,rovinelli_use_1977,nunnally_psychometric_1994,crocker_introduction_2008,miller_measurement_2009, engelhardt_introduction_2009, lindell_establishing_2013,laverty_analysis_2018}. Commonly accepted methods for establishing the validity of an item generally include statistical approaches such as Classical Test Theory and Item Response Theory, though other measures of validity (including through consultation with content experts and the establishment of evidentiary arguments) also exist.

As couplets replace items as the unit of assessment, couplet scores replace item scores in statistical validation procedures. Many of the common statistical validation approaches can be easily adopted to work with couplets instead of items, and, for SPRUCE, this is the focus of a recent paper \cite{geschwind_evidence_2024}. It is also worth noting that the validation of instruments using couplet scoring should primarily focus on ensuring that AO scores from couplets are meaningful measures of those constructs, and that common statistical metrics and thresholds may need to be reevaluated when used with couplet scores. Validation considerations for couplet scoring are discussed further in Sec.~\ref{validation}.

\subsection{Benefits to Users}

This section discusses benefits to users of content RBAIs. By "users", we are referring to the educators who are using these instruments to improve their instruction or to conduct research. 

\subsubsection{Alignment with Instructional Priorities}

Prior to using a content RBAI, instructors and researchers need to ensure the instrument aligns with the content they wish to assess. By having the instrument constructs articulated as AOs (that are used in item development and scoring), direct comparison between the instrument objectives and instructional learning objectives is possible: if an instructor finds that AOs for an assessment match their own learning objectives, then it is likely that assessment will be of use to them. This alignment, known as curricular validity, is ``how well test items represent the objective of the curriculum''~\citep{mcclung_developing_1978}\note{p 87}, and is an important consideration of instructors looking to use RBAIs in their course.

For many instruments, a clear list of instrument constructs is not articulated. With other instruments, the constructs are listed only in academic articles and not presented to implementers alongside the instrument. As such, an implementer would need to either attempt to interpret the intent of the developer and/or review published academic articles detailing instrument development in order to establish curricular validity. With couplet scoring, these objectives are a central part of the instrument reporting, and are straightforward to present to implementers along with the instrument (see Ref~\cite{noauthor_spruce_nodate} for an example).

\subsubsection{Data Yield}

Items that align with more than one AO will have more than one couplet, and, as the couplet is the unit of assessment in couplet scoring, this means that using couplet scoring allows researchers and instructors to get more data from the same number of items, as compared to traditional item scoring. This feature can help to reduce the overall number of items in an instrument, making it easier for students to complete. 

As an example, for SPRUCE, numeric-open-response items allowed us to evaluate students use of significant digits independent of the numeric value they chose to report, and we were able to do so without presenting students with additional items.

Additionally, analysis can be done at the AO level as well as at the item level and the instrument level (see Ref.~\cite{geschwind_evidence_2024} for an example).

\subsubsection{User Experience}
As couplet scoring is essentially a back-end feature, the process of completing an RBAI that uses couplet scoring is virtually indistinguishable from completing an RBAI using traditional item scoring. The only perceivable difference for students might be that the items may be more complex than with traditionally scored RBAIs.

\subsubsection{Reporting Scores by AO}

Central to couplet scoring is the idea that couplets produce scores that are aligned with specific AOs. By reporting scores independently for each instrument AO, and by having the AOs be clear and concise statements about student proficiencies, we argue that the results of an instrument that uses couplet scoring are better contextualized and more actionable for instructors than are single, overall assessment scores. While many instrument reports could be modified to show scores by object/principle/etc., with couplet scoring, producing and presenting AOs scores is straightforward and also in line with the theoretical foundation of couplet scoring.

\section{Developing and implementing an RBAI that uses couplet scoring}
\label{sec implementation}

Now that we have reviewed the concept of a couplet and described many of the benefits of couplet scoring, we now discuss details and limitations of implementing a couplet scoring scheme while designing a physics content RBAI. This section draws on primarily our experience developing SPRUCE, which employed and expanded on the assessment development framework of evidence-centered design (ECD)~\citep{michelle_m_riconscente_evidence-centered_2015}.

The five layers of assessment development as defined by ECD and modified for couplet scoring, are:

\begin{itemize}[itemsep=-.2ex]	
    \item \textit{Domain Analysis}: the collection of information about the topic to be assessed from texts, research literature, interviews with experts, etc.
    
    \item \textit{Domain Model}: the distillation of information from the \textit{domain analysis} into AOs and potential item contexts, including detailing acceptable evidence of proficiencies and the methods for gathering this evidence.
   
    \item \textit{Conceptual Assessment Framework}: the determination of appropriate and desirable assessment features and item formats to support gathering evidence of student proficiencies based on the domain model.
   
    \item \textit{Assessment Implementation}: the writing and piloting of items (and couplets), and the revising of items, AOs, and couplets, to establish evidentiary arguments linking student responses to student reasoning.
    
    \item \textit{Assessment Delivery}: the implementation of the finalized items, scoring scheme, and feedback for implementers.
\end{itemize}

\noindent Many of these layers are similar to steps described in other assessment development frameworks~\citep{laverty_analysis_2018,adams_development_2011}. In the following sections, we discuss these steps and important considerations for assessment developers aiming to implement a couplet-scoring scheme while employing their choice of assessment development frameworks.

\subsection{Developing Assessment Objectives}

The process of collecting information on the topic to be assessed and processing that information into objectives and proto-items (the \textit{domain analysis} and \textit{domain model} in ECD) begins similar to any effort to determine the priorities and objectives of an assessment. Such efforts include consulting the education research literature and commonly used textbooks, identifying and reviewing existing assessments on similar topics, and soliciting priorities and feedback from instructors and other content specialists.

However, as the name ``assessment objective'' suggests, the distillation of this information into constructs should be expressed in terms of desired student proficiency, not just the name of topic to be addressed. For example, SPRUCE contains the AO ``S2 - Identify actions that might improve precision.'' Articulated in this way, AO S2 describes a measurable objective (much like a course learning objective~\citep{anderson_taxonomy_2001}), as opposed to just stating that the instrument assesses the concept of ``precision,'' which is ambiguous as to what specific knowledge or skills around precision will be assessed. As discussed in the previous section, these AOs can have conceptual overlap and need not be wholly independent, since they will be evaluated and reported independent of one another. AOs may be added, removed, split, consolidated, or otherwise refined throughout instrument development, as long as they continue to align with the information gathered in the first stages of development (i.e., in the \textit{domain analysis} for ECD).

\subsection{How to develop items with couplet scoring}

The process of creating items that align with multiple instrument AOs involves many of the steps of traditional item creation. For couplet scoring, these steps are necessarily iterative, which is not always the case for instruments using a traditional scoring scheme: as we saw with item 18 from the FCI, scoring may reveal an issue that requires reevaluation of the AOs. The development of items and a couplet-scoring scheme will likely necessitate expanding, narrowing, splitting, consolidating, or otherwise changing specific AOs, other items that target those AOs will need to be revisited to ensure they still align with the updated AOs. If significant enough modifications are made to items, additional piloting may be appropriate.

Initially, in the first stage of instrument development (the analysis of the topic to be assessed), the assessment priorities of instructors and other experts are being assembled into AOs. At the same time, instrument developers can begin to gather ideas that will inform the creation of specific tasks, as well as logistical considerations that may inform the types of items that are viable and reasonable for these tasks. These steps, for traditionally scored items, are described in many assessment development frameworks, such as in layers one through three of ECD~\citep{michelle_m_riconscente_evidence-centered_2015}.

In the next steps of instrument development, when items are being crafted and before they have been piloted, development for instruments using couplet scoring will differ from more traditional instruments in that the items can reflect a level of complexity representative of instruction, as they can address more than one concept. For selected-response items, initial detractors can be developed by considering the responses that respondents might provide if they were to employ correct reasoning along some of the item's AOs, but incorrect reasoning along other, and the scoring of an item with such distractors should vary between different couplets of the same item.

Once the items have been drafted, the process of item refinement continues much like that of traditionally scored items: an iterative process of piloting the items and refining the item prompts, answer options, and scoring. However, if it is found that a particular couplet is inappropriate or too difficult to assess, it may be possible to remove that particular couplet from the instrument without discarding the entire item. This happened with SPRUCE where, for example, a task asking students to report a best estimate of a value based on a set of repeated measurements dropped a couplet relating to removing outliers, but the item remained in SPRUCE because other couplets for this item were still viable. 

\subsection{Scoring Couplets}

For single-AO items (i.e., items with just one couplet), the process of developing a scoring scheme is much the same as with traditional items, with the added benefit of having an explicit AO guiding scoring to help ensure that the item measures what it was intended to measure.

It can  be tricky initially for content experts, who are used to coordinating many different ideas at once, to look at a multi-AO item and consider how each response relates to only one AO at a time. Such compartmentalization is relatively straightforward for multiple-choice items with AOs that have no conceptual overlap, such as with SPRUCE item 3.3 and the scoring scheme shown in Table~\ref{table 33scoring}, however it can be more challenging for multiple response or coupled multiple-response item formats and AOs that are more closely related. Fortunately, having the instrument and item constructs clearly articulated as objectives provides an easy reference for developers, and the process of scoring by AO can quickly become intuitive. In fact, for SPRUCE, the developers ultimately found the process of scoring by AO made scoring easier for complex items types such as coupled multiple response items, as the AOs themselves productively narrowed the idea-space the developers were considering at any one time.

Additionally, couplet scoring lends itself to an intuitive check of the scoring scheme early on in item development, as demonstrated with the example scoring scheme from item 18 on the FCI discussed in Sec.~\ref{couplet scoring}. This approach involves considering a (temporary) ``collapse'' of the scoring scheme into a partial-credit scheme to gauge possible item responses in terms of the fraction of the total points---across AOs---they would receive: a response aligning with only one of three AOs would receive a third of the total points for the item, for example. Developers can then determine if the relative scores of various answer options seem reasonable, and such checks can help identify potential issues with items, which may be especially useful before piloting with students has occurred.

\subsection{Statistical Validation}
\label{validation}

As discussed in Sec.~\ref{sec affordances}, by replacing items with couplets as the unit of assessment, instruments employing couplet-scoring schemes can use many common statistical validation approaches. However, just as a considering a `total item score' for a multi-AO item can serve as a check to identify large issues in the scoring of couplets for an item, validation metrics that use an overall assessment score can be used as a check to identify large issues  in an instrument. As couplet scoring is designed to score and report proficiencies on an AO-by-AO basis, it may not be reasonable to expect that metrics based on a total score necessarily adhere to the conventional thresholds of single-construct items and instruments. For example, pure guessing on SPRUCE item 3.3 (Fig.~\ref{fig 33}) would result in an average score of 50\% for AO H1 (Table~\ref{table 33scoring}), which is higher than what is generally desired for item scores with $\sim$4 answer choices~\citep{crocker_introduction_2008}. However, it is important to keep in mind that this item also serves as a measure of AO H2, and that the AO score for AO H1 also takes into account other items, contributing to an overall more reliable measure of the AO. 

Additionally, CTT metrics may be calculated using traditional methods by simply replacing items with couplets. The calculations remain the same, although the interpretations may differ slightly. For example, Doran suggests that difficulty should be between 0.30 and 0.90 for each item~\citep{doran_basic_1980}. However, this assumes each item has a four-option multiple choice format, where a difficulty of 0.25 would represent random guessing. Clearly, for some couplets, this assumption no longer holds --- for example, in the previously mentioned SPRUCE item 3.3 example, for AO H1, 50\% is random guessing.

Similarly, Englehardt~\citep{engelhardt_introduction_2009} discusses that difficulty per item should be about 50\% to obtain the best discrimination for each item. Again, this interpretation is not necessarily true when couplet scoring is used and multiple answer options can get full credit. Discrimination should be calculated for each individual couplet, and there may not be the typical correlation with difficulty that one might expect.

Fortunately, the raw numbers for other forms of CTT validation still follow the usual conventions, such as interpretations of discrimination, and reliability in the forms of test-retest stability, as well as internal consistency (Cronbach's alpha).

In addition to performing a CTT validation of the entire test (using statistics such as Cronbach's alpha and Ferguson's delta), as well as on a couplet-by-couplet basis (using statistics such as couplet difficulty and couplet discrimination), some CTT statistics can be applied on an AO basis. For example, one could examine the Cronbach's alpha within a particular AO to determine whether all of the items assigned to that AO are consistent with one another. Additionally, we can calculate difficulty and discrimination at the AO level, answering questions about how difficult certain concepts are for students, as well as how well specific AOs discriminate between high and low performers. Instead of performing CTT on only a couplet and entire assessment level, calculating CTT statistics on AOs allows for an intermediate level of validation.

We applied CTT metrics to SPRUCE data and the results of which are presented in a prior publication~\cite{geschwind_evidence_2024}. In that paper, we showed that this intermediate level of validation is possible and helpful in providing evidence of validity and reliability for the assessment. We also demonstrated that at the couplet, AO, and whole-assessment level that SPRUCE does have evidence of both validity and reliability. In particular, at both the couplet and AO levels, the CTT statistics of discrimination index, difficulty, and Pearson coefficient all showed evidence that the individual couplets, as well as AOs, are valid and reliable. Further, at the entire assessment level, difficulty, Ferguson's delta, Cronbach's alpha, test-retest stability, and split-halves reliability all point towards SPRUCE, as a whole, being valid and reliable. We point readers to this paper to see the details of applying CTT to an assessment scored using the couplet scoring scheme.

\subsection{Instructor Reports}

Instruments that employ couplet scoring will produce a score for each of the instruments AOs, the mean score of several couplets that all target the same AO. These AO scores can be presented to instructors with minimal elaboration, as long as the AOs are clearly written. Thus, AO scores should provide specific and actionable feedback for instructors compared to instruments that use traditional item scores to present instructors with a single number and/or a number for each item. When the instrument is used in a pre-instruction then post-instruction modality, the instrument user can see how proficiency with each AO changes between the beginning and end of their course.

We have previously presented an example figure from an instructor report for SPRUCE that highlights how AO scores are the primary mode of conveying student proficiencies to instructors for an instrument that uses a couplet scoring scheme~\citep{vignal_survey_2023}. Preliminary feedback from instructors on these reports has been overall positive.

\section{Summary}

\label{summary}
In this paper, we discussed a new scoring paradigm, couplet scoring, in which each instrument item is scored potentially multiple times, once for each of the assessment objectives (AOs) that the item aims to measure. We explored how couplet scoring and the use of AOs produce meaningful measures of student proficiency. We then discussed some of the nuances and challenges of implementing a couplet-scoring scheme for a research-based assessment instrument (RBAI), using examples from our work with both developing the Survey of Physics Reasoning on Uncertainty Concepts in Experiments (SPRUCE)~\citep{vignal_survey_2023} and exploring a \textit{post hoc} couplet-scoring scheme for the FCI.

Related future work includes investigating a full \textit{post hoc} couplet scoring scheme for the FCI. Additionally, we believe that elements of this scoring model might be productively employed in formative and summative assessment within a course to support alignment. Research has shown that reporting scores by objective can be  valuable~\citep{mcclung_developing_1978,anderson_taxonomy_2001,mazur_farewell_2009}. However, it is also worth noting that many of the steps of RBAI develop outlined in this paper are not reasonable for a course assessment, and thus we do not claim that all of the benefits of couplet scoring can be realized for a course test.

\section{Acknowledgements}
This work was supported by NSF DUE 1914840, DUE 1913698, and  PHY 2317149. We would like to thank Bethany Wilcox and Katie Rainey for their contributions regarding assessment objectives, and Rachel Henderson for her insights regarding how couplet scoring might be used with statistical validation approaches.

\bibliography{main.bib} 

\end{document}